\documentclass{PoS}

\title{PDFs for nuclear targets}

\ShortTitle{PDFs for nuclear targets}

\author{\speaker{Karol Kova\v{r}\'{\i}k}\\%
       Laboratoire de Physique Subatomique et de Cosmologie (LPSC), Grenoble\\
       E-mail: \email{kovarik@lpsc.in2p3.fr}}

\abstract{Understanding nuclear effects in parton distribution functions (PDF) is 
		an essential component needed to determine the strange and anti-strange quark 
		contributions in the proton. In addition Nuclear Parton Distribution Functions (NPDF) 
		are critically important for any collider experiment with nuclei (e.g. RHIC, ALICE). 
		Here two next-to-leading order $\chi^{2}$-analyses of NPDF are presented. The first uses neutral current charged-lepton $(\ell^{\pm}A)$ Deeply
		Inelastic Scattering (DIS) and Drell-Yan data for several nuclear
		targets and the second uses neutrino-nucleon DIS data. We compare the nuclear corrections factors
		($F_{2}^{Fe}/F_{2}^{D}$) for the charged-lepton data with other
		results from the literature. In particular, we compare and contrast
		fits based upon the charged-lepton DIS data with those using neutrino-nucleon
		DIS data.}

\FullConference{XVIII International Workshop on Deep-Inelastic Scattering and Related Subjects\\
		 April 19 -23, 2010\\
		 Convitto della Calza, Firenze, Italy}

\begin{document}

\section{Introduction}
Parton distribution functions (PDFs) are extremely important in high energy physics as they are needed for the computation
of reactions involving hadrons. For this reason
various groups present and update precise global analyses of PDFs for protons 
\cite{Ball:2009mk,Martin:2009iq,Nadolsky:2008zw,JimenezDelgado:2008hf}
and nuclei \cite{Hirai:2007sx,Eskola:2009uj,deFlorian:2003qf}.
\newline %
PDFs are non-perturbative objects determined by experimental input and the data come from different processes such as the Deep Inelastic Scattering (DIS), 
Drell-Yan (DY) and jet production. A lot of data is extracted from free protons but there is a large share which comes from analysis of data on nuclear targets.
Most prominently, the neutrino DIS on heavy nuclei is proving to be very important for precise determination of the flavor components of the PDFs and it gives the most precise information on the strange quark PDF. The knowledge of the strange quark PDF has an impact on the precision of $W$ and $Z$ boson measurements at the LHC. Data taken on nuclear targets are included in the proton analysis using the nuclear correction factors which are very often based on a specific model \cite{Kulagin:2004ie}. The other option is to use the data on nuclear targets and to extract from them the nuclear parton distribution functions (NPDFs) in order to construct the nuclear correction factors based on experimental input. NPDFs are also used in predictions for collisions of nuclei at RHIC or at the LHC. 
\newline %
Here, we present a new framework for a global analysis of nuclear parton distribution functions at next-to-leading-order (NLO). Then we use it to analyze the apparent differences between nuclear correction factors ($F_{2}^{Fe}/F_{2}^{Fe,0}$) coming from charged lepton data and from neutrino DIS data.
\section{NPDF global analysis framework}
\begin{figure}[t]
\begin{picture}(500,140)(0,0) 
 \put(0,10){\includegraphics[width=0.47\textwidth]{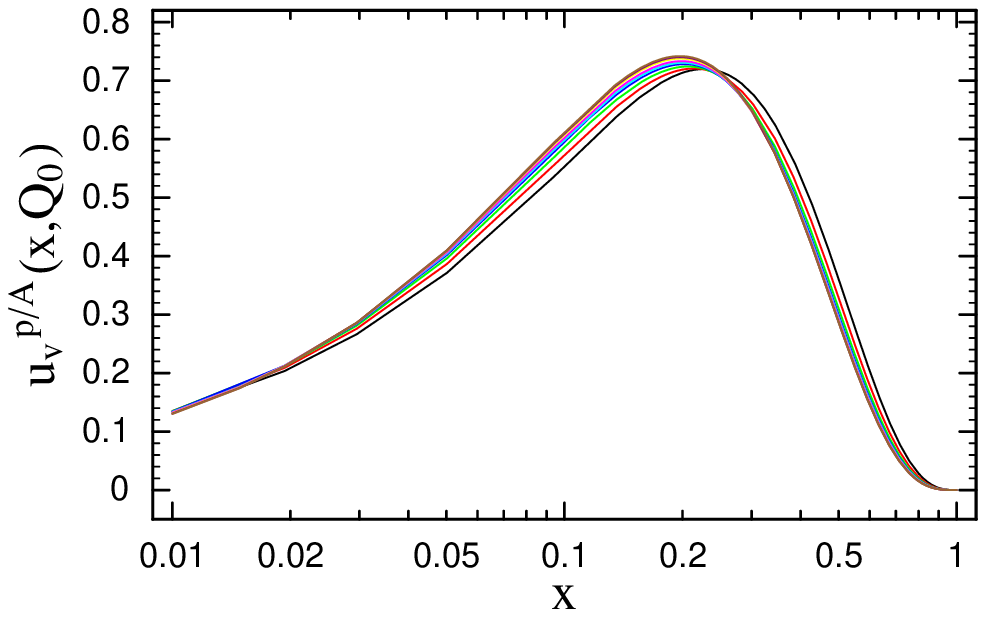}}
\put(215,10){\includegraphics[width=0.47\textwidth]{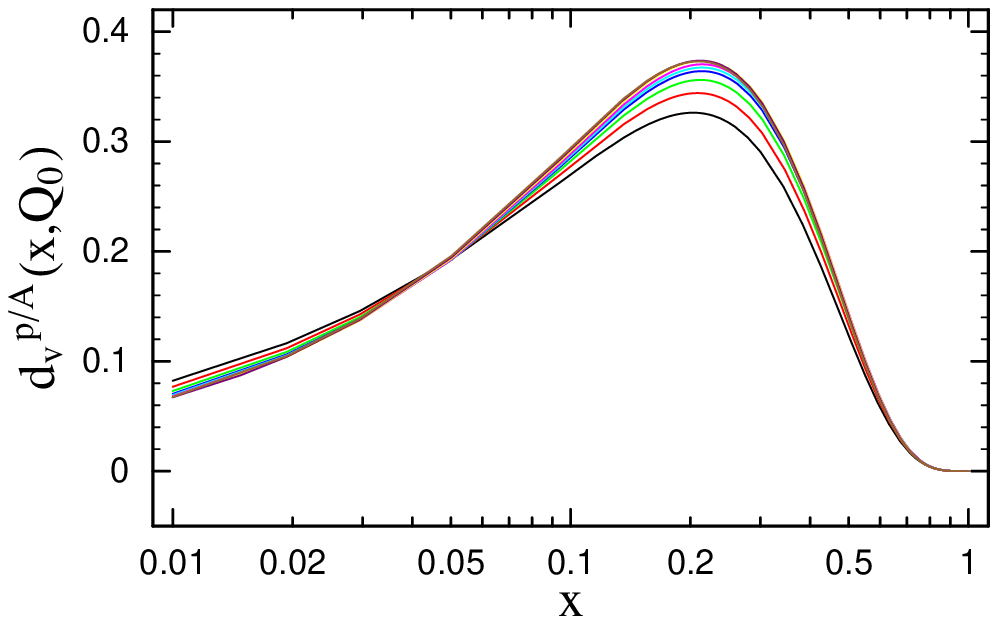}}
\put(109,0){$(a)$} \put(325,0){$(b)$} \end{picture}
\begin{picture}(500,150)(0,0) 
 \put(0,10){\includegraphics[width=0.47\textwidth]{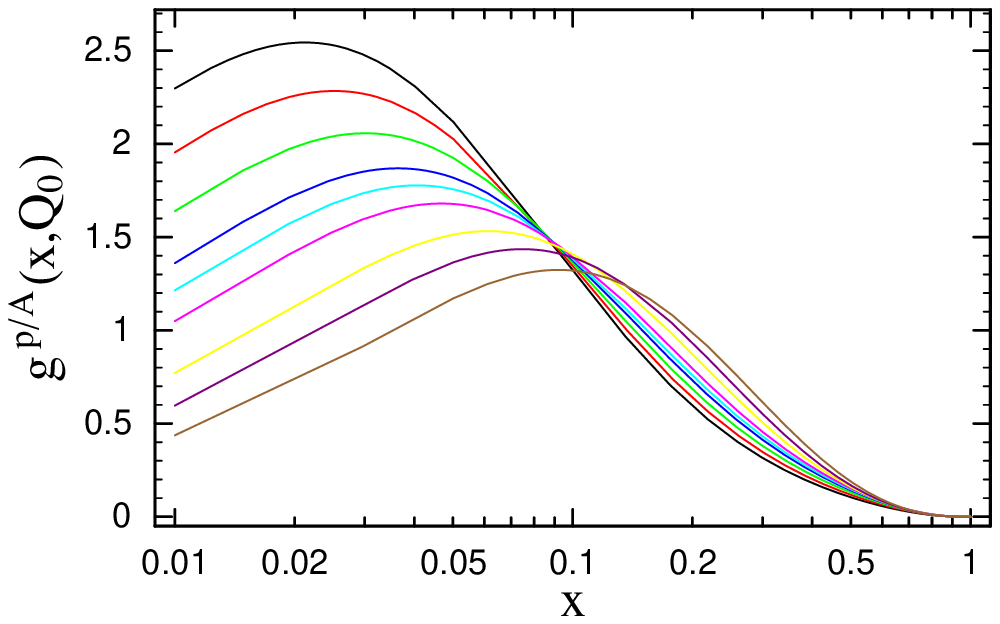}}
\put(215,10){\includegraphics[width=0.47\textwidth]{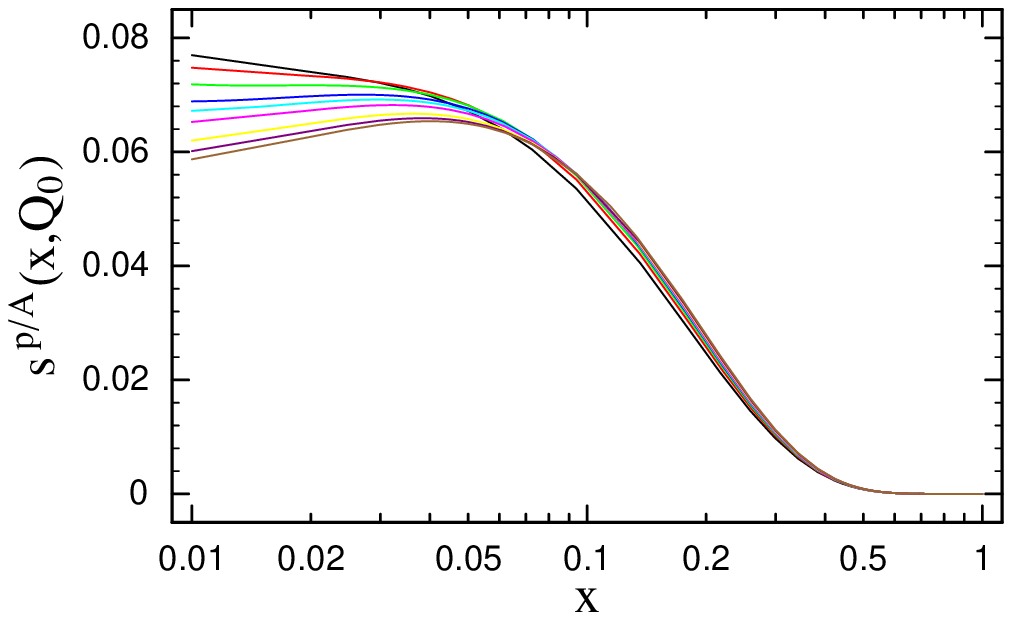}}
\put(109,0){$(c)$} \put(325,0){$(d)$} \end{picture}
\caption{We display the (a) $x\,u_v(x)$, (b) $x\,d_v(x)$, (c) $x\,g(x)$ and (d) $x\,s(x)$, PDFs for a selection of nuclear
$A$ values ranging from $A=\{1,207\}$. We choose $Q_{0}=1.3\,{\rm GeV}$.
The different curves depict the PDFs of nuclei with the following
atomic numbers (from top to bottom in (c) at $x=0.01$) $A=1,2,4,8,20,54,$
and $207$.}\label{Fig:npdf}
\end{figure}

We introduce the global analysis framework and the analysis of NPDFs using the charged lepton DIS and Drell-Yan data for a variety of nuclear targets. The analysis is performed based on the same principle as the proton analysis of \cite{Pumplin:2002vw}. The input distributions are parameterized as
\begin{eqnarray}
x\, f_{k}(x,Q_{0}) & = & c_{0}x^{c_{1}}(1-x)^{c_{2}}e^{c_{3}x}(1+e^{c_{4}}x)^{c_{5}}\qquad\quad k=u_{v},d_{v},g,\bar{u}+\bar{d},s,\bar{s}\,,\label{eq:input}\\
\bar{d}(x,Q_{0})/\bar{u}(x,Q_{0}) & = & c_{0}x^{c_{1}}(1-x)^{c_{2}}+(1+c_{3}x)(1-x)^{c_{4}}\,,\nonumber 
\end{eqnarray}
at the scale $Q_{0}=1.3$~GeV. The different nuclear target materials are treated by introducing
a nuclear $A$-dependence in the $c_{k}$ coefficients: 
\begin{equation}
c_{k}\to c_{k}(A)\equiv c_{k,0}+c_{k,1}\left(1-A^{-c_{k,2}}\right),\quad k=\{1,\ldots,5\}\,.\label{eq:Adep}
\end{equation}
The advantage of this construction is that in the limit $A\to1$ we recover the original proton parameterization
with $c_{k,0}$ as the coefficients. Using this framework, we construct a global fit to the charged lepton DIS data and Drell-Yan data (for details see \cite{Schienbein:2009kk}). The coefficients $c_{k,0}$ were based on the results of the proton global fit presented in \cite{Owens:2007kp} where the influence of nuclear targets on proton PDFs was minimal. In the analysis, we have applied standard kinematic cuts of $Q_{cut}=2.0$~GeV, and $W_{cut}=3.5$~GeV. 
\begin{figure}[t]
\begin{picture}(500,140)(0,0) 
 \put(0,10){\includegraphics[width=0.47\textwidth]{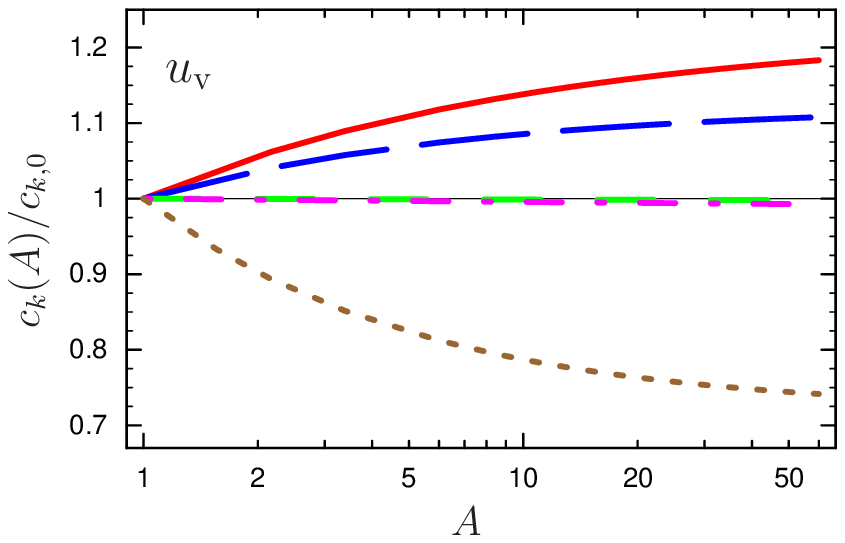}}
\put(215,10){\includegraphics[width=0.47\textwidth]{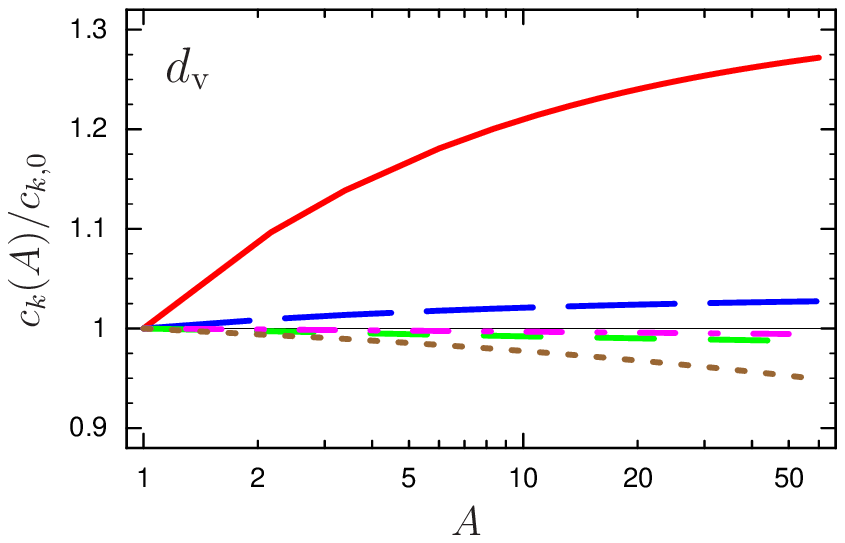}}
\put(107,0){$(a)$} \put(321,0){$(b)$} \end{picture}
\caption{The $A$-dependent coefficients $c_{k}(A)$, $k=\{1,5\}$,
for the up-valence (a) and down-valence PDF (b) as a function
of the nuclear $A$. The different coefficients $c_{k}(A)$
correspond to following lines: $c_{1}$- solid (red) line, $c_{2}$-
long dashed (blue) line, $c_{3}$- dashed (green) line, $c_{4}$-
dash-dotted (magenta) line, $c_{5}$- dotted (brown) line.}\label{Fig:ck}
\end{figure}
Performing the global fit to the data (708 data points after the cuts are applied), we obtain an overall $\chi^{2}/{\rm dof}$ of 0.946 with about 32 free parameters. The results of the global fit, the coefficients $c_{k,1}$ and $c_{k,2}$, give the $A$-dependence of the generalized coefficients $c_k (A)$ (see Fig.~\ref{Fig:ck}) and these coefficients determine the parton distribution functions for bound partons inside a nucleus (see Fig.~\ref{Fig:npdf}). The nuclear effects are typically given in terms of nuclear correction factors $R[F_2^{l^\pm Fe}]\equiv F_{2}^{Fe}/F_{2}^{Fe,0}$ and we show the nuclear correction factors resulting from our fit to charge lepton data in Fig.~\ref{Fig:f2_1}. The nuclear structure functions $F_{2}^{Fe}$ and $F_{2}^{Fe,0}$ are both defined according to
\begin{equation}
F_{i}^{(A,Z)}(x,Q)=\frac{Z}{A}\ F_{i}^{p/A}(x,Q)+\frac{(A-Z)}{A}\ F_{i}^{n/A}(x,Q)\,.\label{eq:sfs}
\end{equation}
These structure functions can be computed at next-to-leading order
as convolutions of the (nuclear) PDFs with the conventional Wilson coefficients,
\textit{i.e.}, generically 
\begin{equation}
F_{i}^{(A,Z)}(x,Q)=\sum_{k}C_{ik}\otimes f_{k}^{(A,Z)}\,.\label{eq:f2ck}
\end{equation}
The difference between $F_{2}^{Fe}$ and $F_{2}^{Fe,0}$ is in using different PDFs ($F_{2}^{Fe}$ uses nuclear PDFs and $F_{2}^{Fe,0}$ uses proton PDFs) in Eq.~(\ref{eq:f2ck}).
\section{Nuclear correction factors from neutrino DIS}
\begin{figure}[t]
\begin{picture}(500,145)(0,0) 
 \put(0,0){\includegraphics[width=0.51\textwidth]{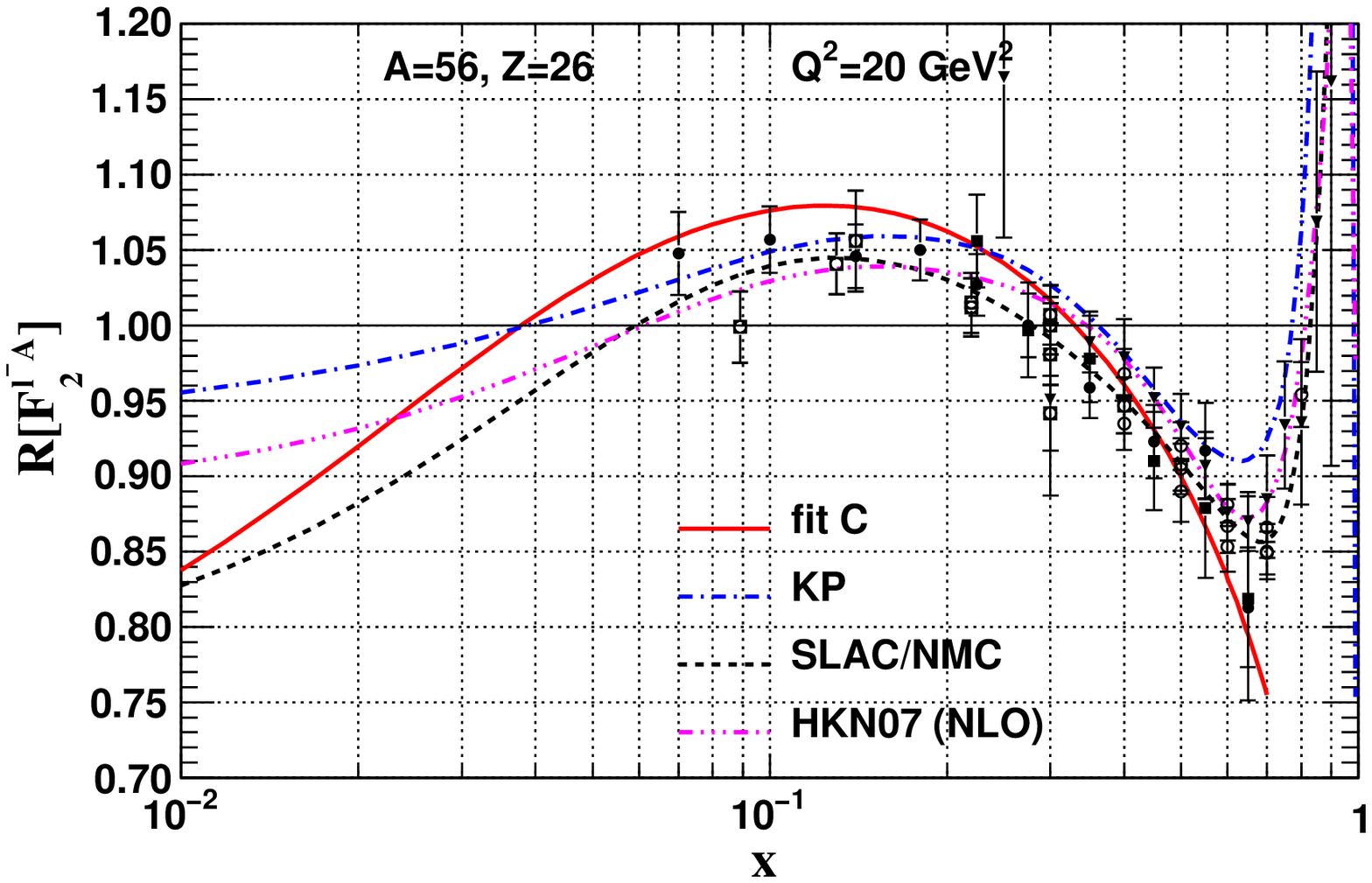}}
\put(215,0){\includegraphics[width=0.51\textwidth]{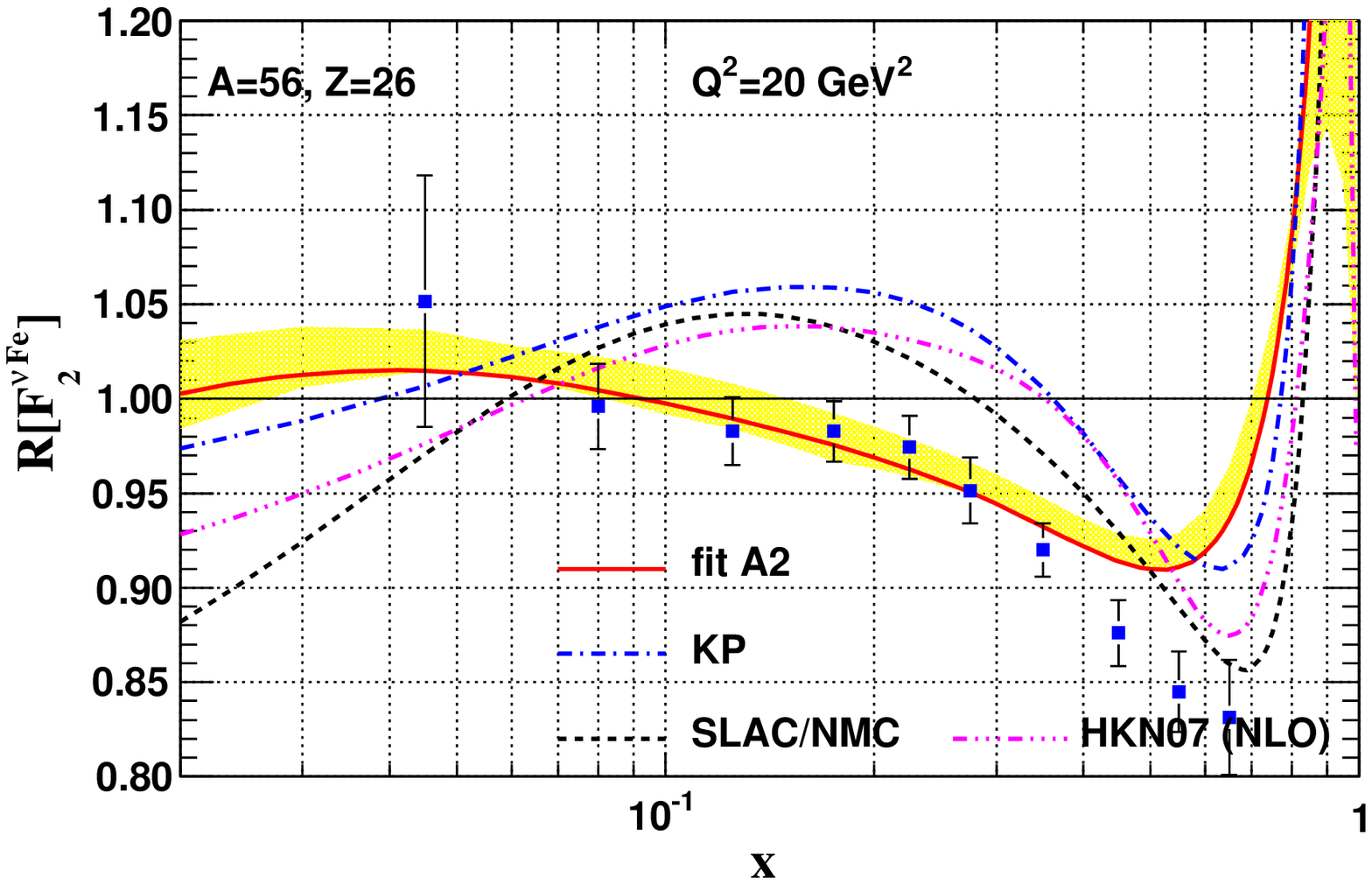}}
\put(110,0){$(a)$} \put(320,0){$(b)$} \end{picture}
\caption{The nuclear correction factor as
a function of $x$ for $Q^{2}=20\,{\rm GeV}^{2}$. Figure-(a) shows
the fit (fit C) using charged-lepton--nucleus DIS and
DY data whereas Figure-(b) shows the fit using neutrino-nucleus
data (fit A2 from Ref.~\cite{Schienbein:2007fs}). Both fits are
compared with the SLAC/NMC parameterization, as well as fits from
Kulagin-Petti (KP) (Ref.~\cite{Kulagin:2004ie})
and Hirai et \textit{al.} (HKN07), (Ref.~\cite{Hirai:2007sx}).}\label{Fig:f2_1}
\end{figure}

\begin{figure}[t]
\begin{picture}(500,145)(0,0) 
 \put(0,0){\includegraphics[width=0.51\textwidth]{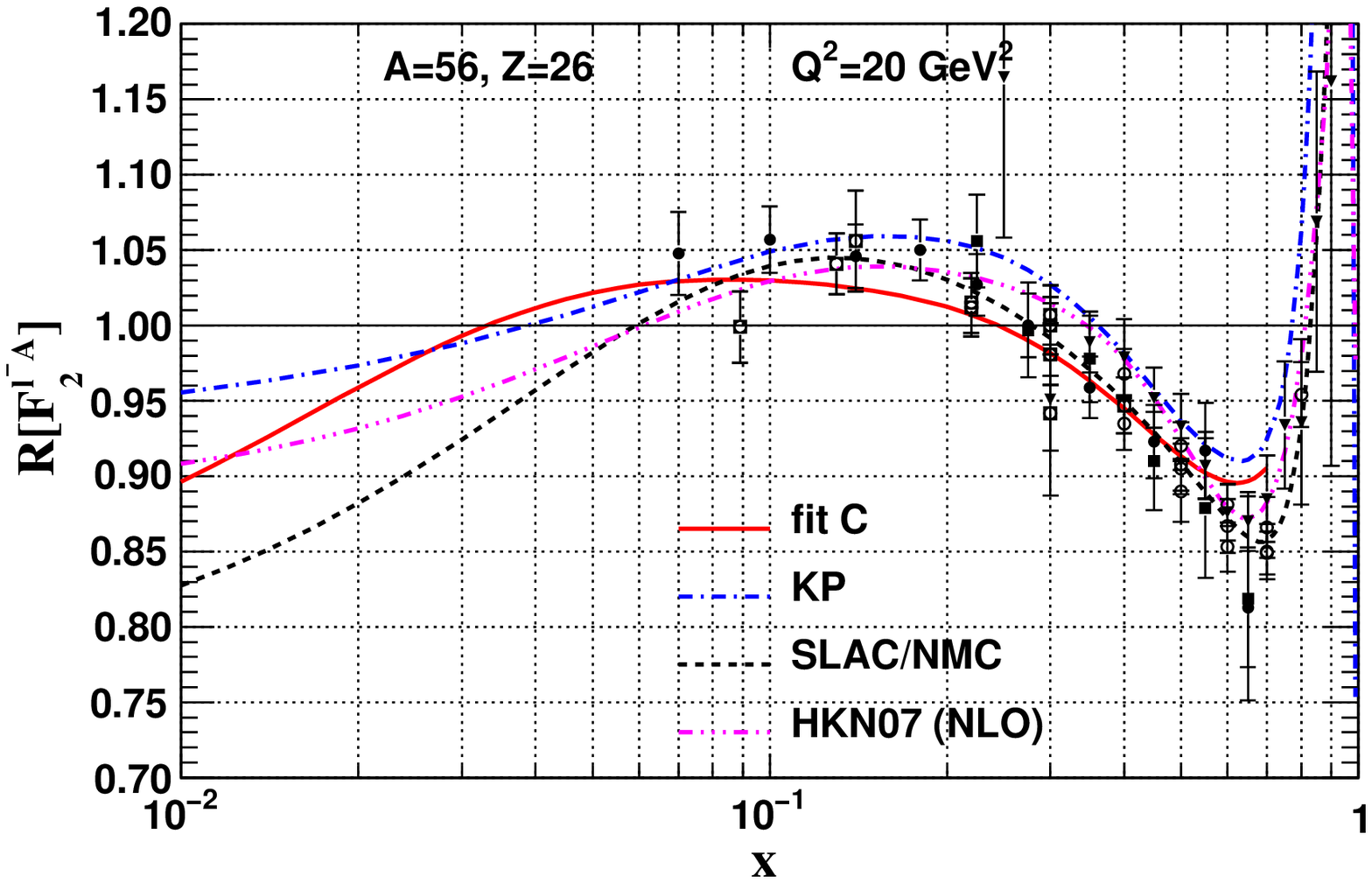}}
\put(215,0){\includegraphics[width=0.51\textwidth]{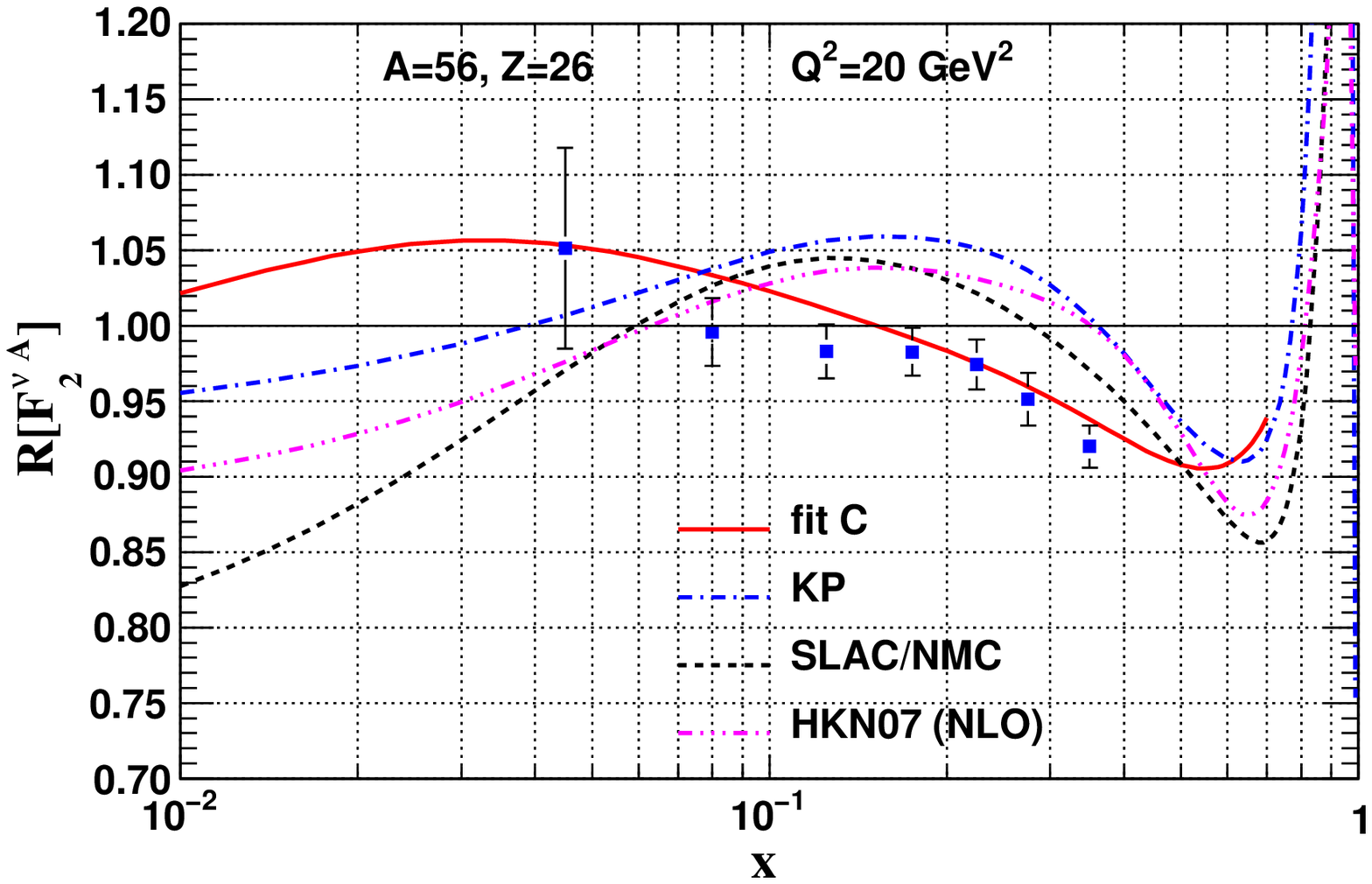}}
\put(110,0){$(a)$} \put(320,0){$(b)$} \end{picture}
\caption{As above but here Figure-(a) shows a compromise fit 
using all charged lepton and neutrino data with neutrino cross-section data weighted down by a factor of $1/2$ whereas Figure-(b) shows the fit using only neutrino and anti-neutrino DIS data.\label{Fig:f2_2}}
\end{figure}

The result of an analysis of NuTeV neutrino DIS cross-section data performed in \cite{Schienbein:2007fs} showed a deviation from the standard result of the analysis of charged lepton DIS and DY data. This can be clearly seen when comparing the different nuclear correction factors in Fig.~\ref{Fig:f2_1}. As the different nuclear correction factors were not obtained in completely identical frameworks, we first used the NPDF framework introduced in the previous section and in \cite{Schienbein:2009kk} to re-analyze the NuTeV neutrino DIS data. 
Using the same kinematic cuts, we obtain a fit to 2310 neutrino and anti-neutrino cross-section and di-muon data points. The result of the fit (see Fig.~\ref{Fig:f2_2}b) is in agreement with the previous analysis and confirms the difference between the nuclear correction factors mainly in the intermediate $x$-region. The obvious difference poses a question if a compromise nuclear correction factor can be found which would accommodate both the charged lepton and neutrino data. In order to construct a compromise fit we use all data used for the charged lepton fit (708 data points), NuTeV and Chorus neutrino and anti-neutrino cross-section data and NuTeV and CCFR di-muon data (3134 data points). To avoid the neutrino data to dominate the analysis, we apply a weight factor $1/2$ to the $\chi^2$ coming from the neutrino and anti-neutrino cross-section data. The result of such a compromise fit with the weight $1/2$ is shown in Fig.~\ref{Fig:f2_2}a. Although the nuclear correction factor from the compromise fit seems to be compatible with the charged lepton data, to draw a firm conclusion a further investigation is necessary. A detailed analysis is postponed to the next publication (see also analysis in \cite{Paukkunen:2010hb}).
\section{Conclusions}
We presented a framework for a global analysis of NPDFs at next-to-leading order QCD closely linked to a proton analysis. 
We used this framework to analyze the discrepancies between the nuclear correction factors stemming from the analysis of charged lepton DIS and DY data 
and the ones coming from neutrino DIS data. We confirm the differences found in a previous analysis and we presented preliminary results 
on compromise fit combining charged lepton and neutrino data. A much more detailed analysis is postponed to a later publication.


\begin{thebibliography}{99}
\bibitem{Ball:2009mk}
R.~D.~Ball et~al.  [The NNPDF Collaboration],
{\em Nucl. Phys.},  B823:195--233, 2009, 0906.1958.
%
\bibitem{Martin:2009iq}
A.~D.~Martin, W.~J.~Stirling, R.~S.~Thorne and G.~Watt,
{\em Eur. Phys. J.}, C63:189--285, 2009, 0901.0002.
%
\bibitem{Nadolsky:2008zw}
Pavel~M. Nadolsky et~al.
\newblock {\em Phys. Rev.}, D78:013004, 2008, 0802.0007.
%
\bibitem{JimenezDelgado:2008hf}
P.~Jimenez-Delgado and E.~Reya.
\newblock {\em Phys. Rev.}, D79:074023, 2009, 0810.4274.
%
\bibitem{Hirai:2007sx}
M.~Hirai, S.~Kumano and T.~H.~Nagai,
{\em Phys. Rev.}, C76:065207, 2007, 0709.3038.
%
\bibitem{Eskola:2009uj}
K.~J. Eskola, H.~Paukkunen, and C.~A. Salgado.
\newblock {\em JHEP}, 04:065, 2009, 0902.4154.
%
\bibitem{deFlorian:2003qf}
D.~de~Florian and R.~Sassot.
\newblock {\em Phys. Rev.}, D69:074028, 2004, hep-ph/0311227.
%
\bibitem{Kulagin:2004ie}
S.~A. Kulagin and R.~Petti.
\newblock {\em Nucl. Phys.}, A765:126--187, 2006, hep-ph/0412425.
%
\bibitem{Pumplin:2002vw}
J.~Pumplin et~al.
\newblock {\em JHEP}, 07:012, 2002, hep-ph/0201195.
%
\bibitem{Owens:2007kp}
J.~F. Owens et~al.
\newblock 2007, hep-ph/0702159.
%
\bibitem{Schienbein:2009kk}
  I.~Schienbein, J.~Y.~Yu, K.~Kova\v{r}\'{\i}k, C.~Keppel, J.~G.~Morf\'{\i}n, F.~Olness and J.~F.~Owens,
\newblock  {\em Phys. Rev.}, D80:094004, 2009, 0907.2357.
%
\bibitem{Schienbein:2007fs}
  I.~Schienbein et~al.
  {\em Phys. Rev.}, D77:054013, 2008, 0710.4897.
%
\bibitem{Paukkunen:2010hb}
  H.~Paukkunen and C.~A.~Salgado,
  arXiv:1004.3140.
%
\end{thebibliography}
\end{document}